\documentstyle[emulateapj,epsfig]{article}

\newcommand{\Pmode}{P_{\rm mode}}
\newcommand{\Pspin}{P_{\rm spin}}
\newcommand{\msun}{M_\odot}
\newcommand{\tauth}{\tau_{\rm th}}
\newcommand{\tauion}{\tau_{\rm ioniz}}
\newcommand{\kms}{\ {\rm km\ s^{-1}}}
\newcommand{\vrot}{v_{\rm rot}}
\def\figcaption[#1]#2{\begin{figure}%
        \begin{center} %
        \epsfig{file=#1}%
        \end{center} %
	\vskip 1in %
        \def\@captype{figure} %
        \caption{#2}%
        \end{figure}}

\lefthead{Ushomirsky \& Bildsten}
\righthead{Rapid Rotation and g-Modes in B stars}
\submitted{Accepted to ApJ Letters, February 18, 1998}

\begin{document}
\title{Rapid Rotation and Nonradial Pulsations: $\kappa$-Mechanism
Excitation of g-Modes in B Stars}

\author{Greg Ushomirsky and Lars Bildsten} 

\affil{Department of Physics and Department of Astronomy, University
of California, Berkeley, CA 94720}

\begin{abstract}
Several classes of stars (most notably O and B main-sequence stars, as
well as accreting white dwarfs and neutron stars) rotate quite
rapidly, at spin frequencies greater than the typical g-mode
frequencies.  We discuss how rapid rotation modifies the
$\kappa$-mechanism excitation and observability of g-mode
oscillations.  We find that, by affecting the timescale match between
the mode period and the thermal time at the driving zone, rapid
rotation stabilizes some of the g-modes that are excited in a
non-rotating star, and, conversely, excites g-modes that are damped in
absence of rotation.  The fluid velocities and temperature
perturbations are strongly concentrated near the equator for most
g-modes in rapidly rotating stars, which means that a favorable
viewing angle may be required to observe the pulsations.  Moreover,
the stability of modes of the same $l$ but different $m$ is affected
differently by rotation.  We illustrate this by considering g-modes in
Slowly Pulsating B-type stars as a function of the rotation rate.
\end{abstract}

\keywords{stars: early-type --- stars : oscillations --- 
stars : rotation --- stars: variables: other}

\section{Introduction}
\label{sec:intro}

Despite a long-standing interpretation of multiple periodicity in
$\beta$~Cepheids and Slowly Pulsating B-type (SPB,
\cite{Waelkens91}) stars (also referred to as 53 Per stars, after
the prototype discovered by \cite{smith-53per}) in terms of
nonradial pulsations, the details of the excitation mechanism
remained unresolved until quite recently.  The driving of pulsations
in the He$^+$ ionization zone, which gives rise to the classical
Cepheid instability strip, cannot operate in these hot stars.  In a
remarkable example of the interplay between theory and observation,
the updated opacity calculations (\cite{OPAL92}; \cite{OP94}) found a
significant opacity increase in the metal partial ionization zone,
just as suggested by \cite{Simon82} as the explanation for intrinsic
variability in 53 Per stars.  The resulting instability region,
determined by the $\kappa$ mechanism operating in the ``metal opacity
bump'' at $T\approx 1.5\times10^5\ {\rm K}$, incorporates most OB
stars ($M\gtrsim 3M_\odot$) that are in the core hydrogen burning
stage (\cite{Dziembowski-bcep}; Dziembowski, Moskalik, \& Pamyatnykh
1993; \cite{GautschySaio-bstars}).
The most recent excitation calculations for $\beta$~Cepheids and SPBs
(\cite{Pamyatnykh-lanl}) place all known variable stars in these classes
within the theoretical instability region.  However, it is not clear
that {\it all\/} stars in this region are indeed pulsating.

All previous g-mode pulsation calculations have been carried out for
non-rotating or very slowly rotating stars.  However, most OB stars
rotate relatively rapidly and observations indicate that g-mode
pulsations and rotation are somehow related.  As discussed in
\S\ref{sec:freqs}, rotation strongly affects g-mode properties when
$2\Pmode\gtrsim\Pspin$, where $\Pmode$ is the mode period in the {\it
co-rotating} frame, and $\Pspin$ is the spin period.  Rotation would
have to be near breakup to affect the p-mode pulsations in the $\beta$
Cepheids.  However, typical SPB stars ($M\approx 4 M_\odot$, $R\approx
3R_\odot$) exhibit high radial order g-mode pulsations of day-like
periods, so that rotational velocities
\begin{equation}\label{eq:rot-condition}
\vrot\ge 75 \kms \left(\frac{1\ {\rm day}}{\Pmode}\right)
\left(\frac{R}{3 R_\odot}\right),
\end{equation}
will significantly affect the g-modes. The mean rotational velocity
for B stars is $\vrot\approx200\kms$ (\cite{Lang-astro}) and thus cannot be
neglected in pulsation calculations.

One of the hallmarks of the photometrically discovered SPBs is that
they are anomalously slow rotators, with projected rotational
velocities $v \sin i \sim 5 - 40 \kms$ (\cite{Waelkens87}; \cite{Waelkens91}).
While most of the $\sim70$ SPBs discovered by Hipparcos
(\cite{Waelkens98}) are also slow rotators, some have $v\sin i$'s up to
$150\kms$, a regime where rotational effects are very important.
Another correlation between rotation and pulsations is that no SPB
stars were found in the open clusters NGC~3293 and NGC~4755, where the
mean projected velocity of B stars is $v\sin i\approx170 \kms$
(\cite{Balona94b}). Nevertheless, there were many $\beta$~Cep stars
discovered {\it photometrically\/} in the same clusters
(\cite{Balona94a}; \cite{BalonaKoen94}), so this was not due to a selection or
metallicity effect. The fact that most photometrically discovered SPBs
are slow rotators, as well as the absence of SPBs in these two
clusters, led \cite{Balona-review} to conjecture that rapid rotation
suppresses g-modes in these stars.

In this paper we explore the interaction between rapid rotation and
nonradial pulsations.  We begin in \S\ref{sec:freqs} by outlining the
effects of rotation on the adiabatic mode properties (frequencies and
eigenfunctions) and observability of g-modes in rotating B stars.  In
\S\ref{sec:excit}, we discuss the interplay between rapid rotation and
excitation of pulsations by the $\kappa$~mechanism and its effect on
the stability of g-modes in B stars. We close by outlining the future
directions for our work and mention other potential applications of
the rotationally modified g-mode theory we are developing.

\section{Adiabatic g-modes in Rapidly Rotating Stars}
\label{sec:freqs}

We begin by reviewing the theory of adiabatic nonradial pulsations in
rotating stars (see \cite{BUC}, hereafter BUC, for a
detailed discussion and references).  Throughout this paper, we assume
that the star rotates uniformly with spin frequency $\Omega$, much
smaller than the breakup frequency $\Omega_b\approx (GM/R^3)^{1/2}$,
and always work in the co-rotating frame.  The star is nearly
spherical in this limit, and the effects of the centrifugal force
(which are of ${\mathcal O}(\Omega/\Omega_b)^2$) can be neglected.
For our purposes, ``rapid rotation'' is the regime in which
$\omega\lesssim\Omega\ll\Omega_b$, where $\omega$ is the mode
frequency in the co-rotating frame.  In this case, the primary
difference in the momentum equation is the Coriolis force, the effect
of which is of ${\mathcal O}(\Omega/\omega)$, and is always fully
included in our calculations.  Perturbation expansions typically
employed in the slow rotation regime break down when rotation is
rapid.  We use the method developed by BUC which converges
regardless of the rotation rate.  When $\Omega\gtrsim\omega$, the
Coriolis force strongly affects the g-modes. On the other hand,
neither the p-modes nor the radial modes (both of which have
$\omega\ga\Omega_b$) are strongly modified when $\Omega \ll \Omega_b$.

Within the traditional and Cowling approximations (see BUC
and references therein), the adiabatic pulsation equations separate,
i.e. the solutions can be written as $Q(r,\theta,\phi,t )=Q_r(r)
Q_\theta(\theta) e^{im\phi+i\omega t}$.  The frequency in the
observer's frame is then $\omega_I=\omega-m\Omega$.  The Cowling
approximation (i.e. neglecting the perturbation of the gravitational
potential) is justified because we are studying high radial order
g-modes.  The traditional approximation amounts to neglecting the
components of the Coriolis force due to the local horizontal component
of the spin.  For g-modes, this is justified when the scale height $h$
is small compared to the radius, $h/r\ll 1$, and the buoyancy is
strong compared to the radial component of the Coriolis force,
$|N^2|\gg r\Omega\omega/h$, where $N$ is the Brunt-V\"{a}is\"{a}l\"{a}
frequency. Both of these conditions are satisfied in the envelopes of
B stars and allow us to make some progress without resorting to a 2-D
calculation. Extending our calculation into the deep interiors will
force us to drop this approximation, however, since most excitation
and damping is confined to the envelope, the results presented in
\S\ref{sec:excit} are qualitatively correct. In the traditional
approximation, the angular part of the solution obeys Laplace's tidal
equation, $L(Q_\theta)=-\lambda Q_\theta$, where $L$ is an angular
operator that depends only on $m$ and the rotation parameter $q\equiv
2\Omega/\omega$ (BUC). Typical solutions $\lambda(q)$ for
g-modes are shown in Figure~2 of BUC.  The radial
equations are identical to those for a non-rotating star, except with
the spherical harmonic eigenvalue $l(l+1)$ replaced by $\lambda$.
When buoyancy is strong, $N^2\gg\omega^2$, if $\omega_0$ is the
frequency of an $l=1$ mode in a {\it non-rotating\/} star, then the
frequency of the same mode at arbitrary spin is just $\omega =
\omega_0 (\lambda/2)^{1/2}$.  Moreover, when $N^2\gg\omega^2$, the
radial dependence of the eigenfunctions is the same is in the
non-rotating case.

One can view $\lambda^{1/2}$ as the effective transverse wave number
(i.e. $k_{tr}^2=\lambda/R^2$).  For no rotation ($q=0$),
$\lambda=l(l+1)$ and is independent of $m$. For slow rotation ($q\ll
1$), the degeneracy between the different $m$ values is lifted,
leading to the familiar $m$-splitting.  However, for rapid rotation
($q\gg 1$), the modes with $l_\mu\geq 1$ nodes in the angular
eigenfunction have $\lambda\simeq (2l_\mu-1)^2 q^2$.  Hence, when the
star is rapidly rotating, the eigenfrequency in the traditional
approximation,
\begin{equation}\label{eq:omega-scaling}
\omega^2\simeq 2^{1/2}(2l_\mu-1)\Omega\omega_0,
\end{equation}
is roughly independent of $m$, and is just a geometric mean of the
spin frequency $\Omega$ and the non-rotating $l=1$ mode frequency
$\omega_0$ for modes with $l_\mu \geq 1$.  \cite{PP78} first found
this scaling in the WKB limit ($l_\mu\gg 1$), and BUC
extended their result to the observationally relevant low-$l$
modes. On the other hand, $\lambda(q)$ for modes with $l_\mu=0$ zero
crossings in the angular direction (i.e. the modes with $m=-l$) stays
low and approaches $\lambda \simeq m^2$ for $q \gg 1$.

The angular dependence of the eigenfunction for rapid rotation is no
longer just a $Y_{lm}$, however, we label the g-mode at arbitrary
$\Omega$ by the $l$ and $m$ values the same g-mode has when
$\Omega=0$. The fluid displacements and pressure perturbations for the
$l_\mu\geq1$ modes are strongly concentrated towards the equator, with
the bulk of the amplitude between the latitudes $-1/q \leq \cos\theta
\leq 1/q$ (Figure~\ref{fig:color}; see also \cite{LeeSaio90};
BUC; Figure~5 of \cite{tow97}).  Therefore, g-modes in
rapidly rotating stars may be harder to detect because pulsation
amplitudes are appreciable only over a small fraction of the stellar
surface.  How this affects detectability depends on how large the
amplitudes are for the modes excited in the rapidly rotating case
compared to the non-rotating case.

\section{Excitation of Rotationally Modified g-modes}
\label{sec:excit}

The most prominent effect of rapid rotation ($q\gg1$) on g-mode
pulsations is that their periods in the co-rotating frame become
shorter, $\Pmode\propto(\Pspin P_0)^{1/2}$, where $P_0$ is the period
of the mode in a non-rotating star (see
eq. [\ref{eq:omega-scaling}]). This scaling is crucial for
understanding the effect of rapid rotation on mode stability, as we
now show.

The qualitative understanding of the effect of rotation on g-mode
excitation comes from the work integral formulation.
Aizenman \& Cox~(1975) derived the work integral for a star possessing
arbitrary fluid flow in the equilibrium configuration.  Since we
assumed rigid rotation, the background stellar model cannot be in
thermal balance.  However, the thermal imbalance effects are of
${\mathcal O}(\Omega/\Omega_b)^2$ and are negligible compared to the
dominant $\kappa$-mechanism terms (\cite{AizenmanCox75}).  The dominant
contribution to the work integral is the familiar expression,
\begin{equation}
\label{eq:work-int}
W=\int\left(\frac{\Delta
T}{T}\right)\Delta\left(\epsilon-\frac{1}{\rho}\vec{\nabla}\cdot
\vec{F}\right) dM \equiv
\int\frac{dW}{dr}dr,
\end{equation}
where $\Delta T$ is the Lagrangian temperature perturbation
(\cite{Gautschy-review-theory}), and the second equality is obtained by
integrating over the angles.  A mode is linearly unstable if $W>0$, so
that regions of the star with $dW/dr>0$ excite the pulsation. Driving
occurs in regions with $d[\kappa_T+\kappa_\rho/(\Gamma_3-1)]/dr>0$,
where $\kappa_T=(\partial\log\kappa/\partial\log T)_\rho$ and
$\kappa_\rho=(\partial\log\kappa/\partial\log\rho)_T$
(\cite{Gautschy-review-theory}), and competes with damping throughout
the rest of the star, where the above inequality is reversed.

We consider pulsational stability of several g-modes in a $4M_\odot$
main-sequence model ($t\approx 10^8$~years, approximately half of
main-sequence lifetime) with $\log L/L_\odot=2.51$, $\log T_{\rm
eff}=4.13$, and $Z=0.02$, computed using the Warsaw-New Jersey stellar
evolution code provided to us by R. Sienkiewicz.  This model is very
similar to the one used by \cite{Dziembowski-spb}. We use the
quasi-adiabatic approximation, i.e. evaluate $W$ using the adiabatic
eigenfunctions. At some point in the stellar envelope the local
thermal time, $\tauth(r)=\int_r^R c_pTdM/L\approx c_p T\Delta M/L$,
becomes much smaller than the mode period $\Pmode$. These
nonadiabatic parts of the envelope do not contribute to either
driving or damping, so we truncate the work integral at
$\tauth=\Pmode$ to approximate this effect.

In SPB stars, the $\kappa$~mechanism acts in the metal ionization
zone, at $\log T\approx 5.2$.  Figure~\ref{fig:opacity-work}a shows
$\kappa_{T}+\kappa_\rho/(\Gamma_3-1)$ for the $4\msun$ SPB model
described above in the region of the metal opacity bump.
Figures~\ref{fig:opacity-work}b and~\ref{fig:opacity-work}c show the
differential contribution to the work integral, $dW/d\log T$, for
$l=2$, $n=14$ and $l=1$, $n=30$ modes, evaluated using {\it adiabatic\/}
eigenfunctions.  The contribution to the work integral from the inner
part of the ionization zone is positive, while the contribution from
the outer part is negative, a typical behavior for the g-modes in SPB
stars.  As $\Omega$ increases, $\Pmode$ in the co-rotating frame
becomes shorter.  However, as discussed in \S\ref{sec:freqs}, the
radial dependence of the eigenfunctions in the traditional
approximation is the same as in the non-rotating star.  Thus, for
low-$l$ SPB pulsations, {\it the dominant effect of rotation is to
truncate the work integral closer to the stellar surface}.  For the
mode in Figure~\ref{fig:opacity-work}b, the period in a non-rotating
star is $\Pmode=0.675\ {\rm days}$, so the work integral is truncated
at $\log T\approx 5.18$, resulting in an unstable mode, $W>0$ (see
Figure~\ref{fig:period}).  However, as the spin frequency is increased
from $\Omega=0$, the period of the mode decreases, and the location in
the star where the mode goes nonadiabatic, $\Pmode=\tauth$, moves
closer to the stellar surface.  In the non-rotating star, all of the
ionization zone already contributes to the excitation of the mode, so
at moderate rotation, the outer, damping, part of the ionization zone
starts contributing as well.  As shown in Figure~\ref{fig:period}, at
$\Pspin\approx30\ {\rm hours}$ ($\vrot\approx140\kms$), the $m=1$ and
$m=2$ harmonics of this mode become stable.  On the other hand, the
mode in Figure~\ref{fig:opacity-work}c, with a $2.72\ {\rm day}$
period is stable when $\Omega=0$.  However, as rotation becomes rapid,
the adiabatic part of the star encompasses more of the driving part of
the ionization zone.  As shown in Figure~\ref{fig:period}, the $m=1$
harmonic becomes unstable ($W>0$) when $\Pspin\approx70\ {\rm hours}$
($\vrot\approx60\kms$).

As evident from the preceding discussion, if the ionization zone is
completely adiabatic (i.e. if the thermal time at the ionization zone
$\tauion\gg\Pmode$), the driving due to the inner part of the
ionization zone is likely to be cancelled by the damping due to the
outer part. On the other hand, if the ionization zone is completely
nonadiabatic (i.e. if $\tauion\ll\Pmode$), then it does not
contribute to either driving or damping.  Thus, to get the optimal
driving contribution from the metal opacity bump in SPBs, {\it the
period of the mode in the co-rotating frame should be comparable to
the thermal time in the ionization zone}, i.e. $\tauion\sim\Pmode$
(\cite{GautschySaio-bstars}).

In close agreement with the calculations of
\cite{Dziembowski-spb}, we find that the $n=14$--$23$ modes with
$l=1$ and $n=14$--$29$, $l=2$ modes are excited in the non-rotating
$4M_\odot$ SPB model.  Rapid rotation damps the modes with $l=1$,
$n=14$--$16$ and $l=2$, $n=14$--$16$, and excites the $l=1$,
$n=24$--$39$ and $l=2$, $n=30$--$40$ modes. On the other hand, for
this model, the modes with $n\approx 20$, in the middle of the excited
range, can be stabilized only if rotation is near breakup, or cannot
be stabilized at all.  This behavior can be understood in terms of the
normalized growth rate of the mode, $\eta=W/\int|dW/dr|dr$
(\cite{ste78}), which measures the robustness of mode excitation.  The
modes with $n\approx 20$ have $\eta\approx 0.1$--$0.2$, while the ones
with $n\approx 15\ {\rm or\ }30$, at the boundaries of the excited
range, have $\eta\approx 0.01$--$0.05$ (Dziembowski et~al. 1993).
This means that if rotation is to stabilize the modes with $n\approx
20$, it has to change $W$ by about $15$--$20\%$, which is difficult to
achieve within our current approximations.  On the other hand, for
$n\approx 15\ {\rm or\ }30$, a few percent change in $W$ can change
the character of the mode, and is easily accomplished by rotation.

The modes with $m\ge 0$ are more strongly affected by rotation than
the $m<0$ modes, because the former reach larger $\lambda$ (shorter
period) at smaller $q$ (slower rotation).  This is an interesting
correlation that may be confirmed observationally.  The $l=1$, $m=-1$
and $l=2$, $m=-2$ modes (not shown in Figure~\ref{fig:period}) are not
strongly affected by rotation.  The value of $\lambda$ for these modes
only changes from $2$ to $1$ and $6$ to $4$, respectively, as rotation
becomes rapid, and thus the periods and stability of these modes are
not significantly modified.

\section{Conclusions}
\label{sec:conclusions}

Though we have made progress in understanding the interaction between
rotation and pulsations, we have yet to unambiguously explain the
preponderance of slow rotators among the field SPB stars.  Rotational
modification of the mode period can destroy the timescale match
$\Pmode\sim\tauion$, and stabilize a g-mode that is excited in a
non-rotating SPB model.  Conversely, if the thermal time at the
driving zone $\tauion$ does not match the non-rotating mode period,
the mode can be ``spun up'' and become excited if the star is rotating
rapidly.  The periods of excited modes in the co-rotating frame are
largely determined by the structure of the star, in order to satisfy
$\Pmode\sim\tauion$.  However, the radial orders $n$ of the excited
g-modes depend on rotation.  In order to maintain the timescale match
with increasing $\Omega$, one needs to increase $n$ (note that the
co-rotating frame periods and stability of g-modes with $m=-l$ is not
significantly affected by rotation).  However, going to a larger $n$
may mean more radiative damping in the stellar interior, especially in
the $\mu$-gradient zone.  In addition, we find that different $m$
harmonics of the same mode are affected to various degrees by
rotation, which may have consequences for line profile observations
capable of determining the $l$ and $m$ values of the pulsation modes.

 Rotation does not significantly affect the g-modes in B stars until
the spin period is comparable to the mode period. Rotation this rapid
confines most g-modes to the equatorial region, requiring a favorable
viewing geometry for detection, and can destroy the timescale match
needed for mode excitation.  In this sense, the correlation between
slow rotation and SPB pulsations may be an observational selection
effect after all, since the g-modes in rapid rotators, even if
present, only cover a small fraction of the stellar surface.  In order
to differentiate between these two effects, one needs information
about the inclination angle $i$, as well as the amplitude of
pulsations.  Though, as explained in \S\ref{sec:freqs}, rotation is
never ``rapid'' (in the sense $2\Omega\gtrsim\omega$) for p-mode
pulsations, \cite{LeeBaraffe95} have shown cases where the low-order
rotational perturbations can affect p-mode stability in $\beta$
Cep-like stars rotating close to breakup.

Our work may also be applicable to other classes of variable stars,
most notably the $\gamma$~Doradus stars.  These stars, named after the
prototype discovered by \cite{cou63}, are early F-type main-sequence
stars that lie at or below the cool edge of the classical Cepheid
instability strip, and show variability with periods of around a day
(\cite{bre95}).  It is currently believed that the variability is due to
g-modes of high radial order (\cite{kri97}).  For an F-type star with
$R\sim 1.5 R_\odot$, rotational velocities that would affect g-modes
with one-day period are $\vrot\ge 40\kms$ (see
eq.~[\ref{eq:rot-condition}]). Typical rotational velocities for early
F stars are $\vrot\approx100\kms$ (\cite{Lang-astro}).  Therefore, if
g-modes are indeed responsible for $\gamma$~Dor variability, rotation
will have a significant effect on the mode properties and stability.

\acknowledgements G.U. acknowledges the Fannie and John Hertz
Foundation for fellowship support, and L.B. acknowledges support from
the Alfred P. Sloan Foundation.  This research was supported by a
Hellman Family Faculty Fund award to L.B.

\newpage

\figcaption[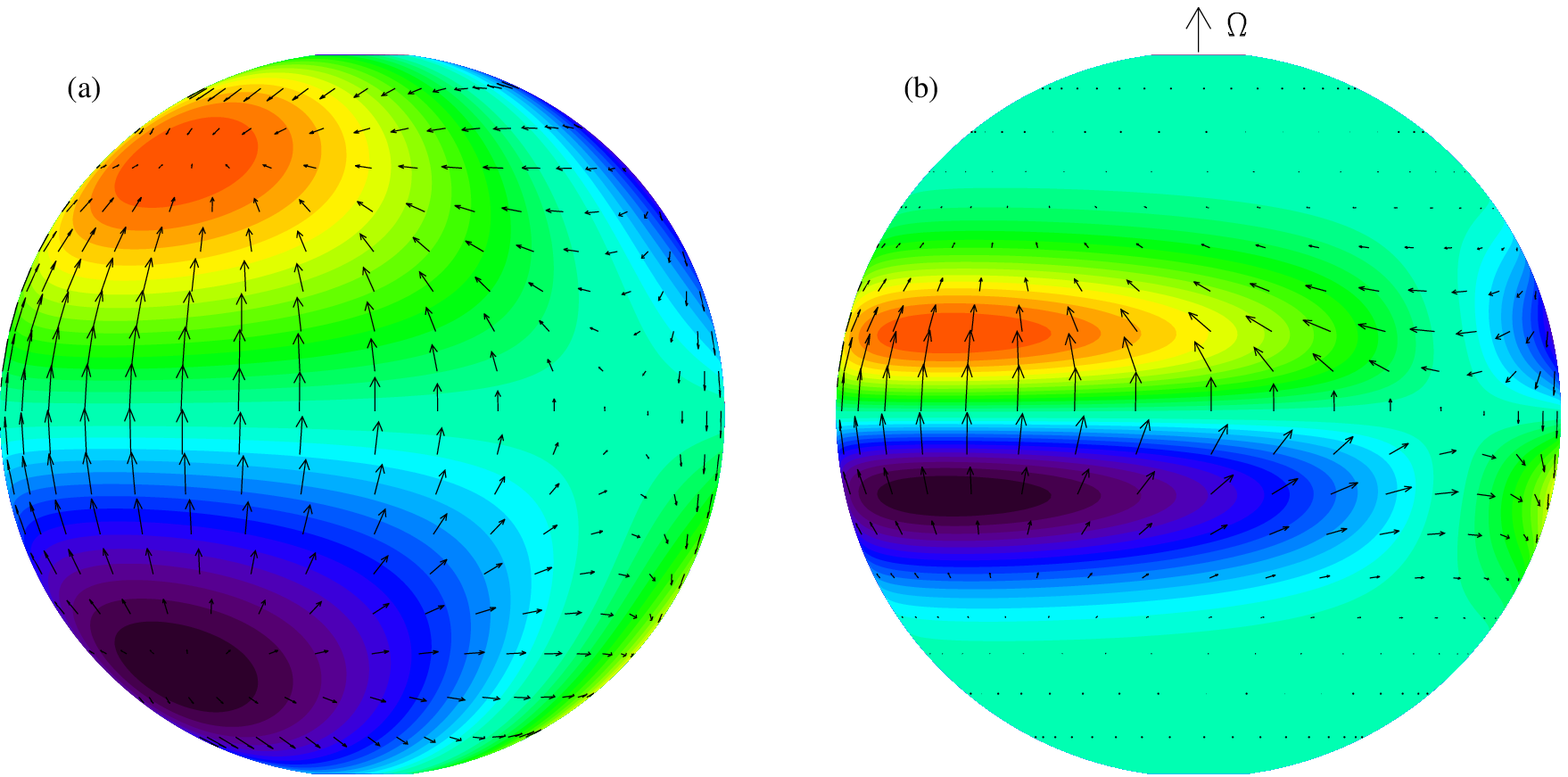]{
\label{fig:color}
(a) Adiabatic Eulerian pressure perturbations (colored contours) and
fluid displacements (black arrows) for an $l=2$, $m=-1$ g-mode in a
non-rotating star.  Normalization is arbitrary, but orange/yellow
(purple/blue) contours denote positive (negative) pressure
perturbations. (b)  Same as (a), but for an $l=2$, $m=-1$ g-mode in a
rapidly rotating star, with a spin period $\Pspin=0.5 \Pmode$.  Note
how pressure perturbations and fluid displacements are strongly
concentrated towards the equator.}

\figcaption[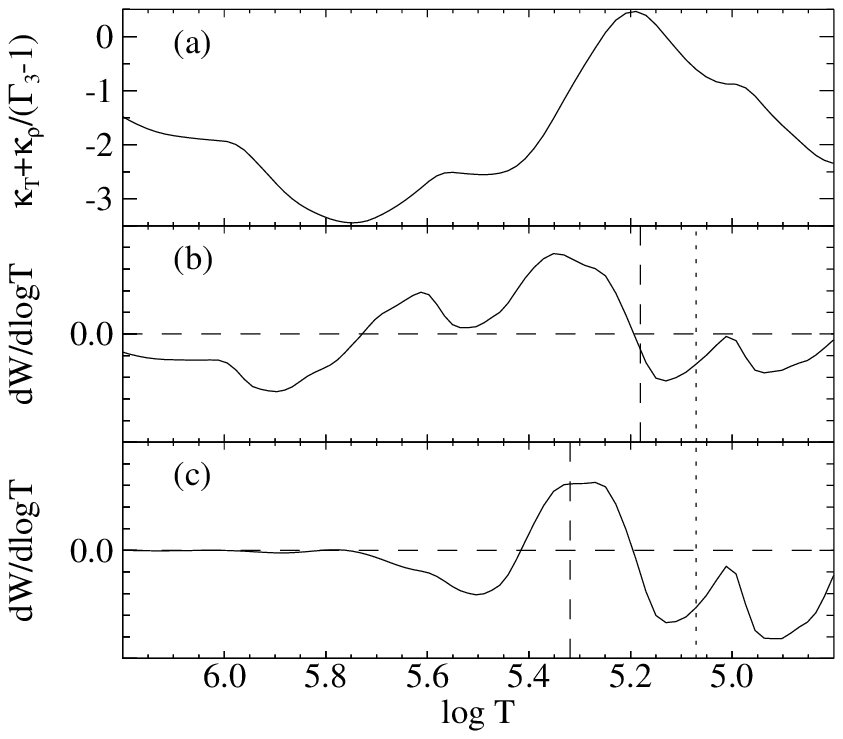]{
\label{fig:opacity-work} 
(a) The opacity derivative $\kappa_{T}+\kappa_{\rho}/(\Gamma_3-1)$ in
a $4M_\odot$ SPB model. (b) Differential work integral, $dW/d\log T$,
evaluated using {\it adiabatic\/} eigenfunctions, for an $l=2$, $n=14$
mode. Dashed and dotted vertical lines show where $\tauth$ is equal to
$\Pmode=0.675\ {\rm days}$, corresponding to the $l=2$ mode period in
the non-rotating star, and the breakup period, $P_b\approx 8.6$~hours,
respectively. In the non-rotating case, $dW/d\log T$ computed using
{\it nonadiabatic\/} eigenfunctions would decay to zero for
temperatures lower than that indicated by the dashed lines. We
approximate this effect by truncating the work integral as described
in \S\ref{sec:excit}.  (c) Same as (b), but for an $l=1$, $n=30$ mode
with $\Pmode=2.72\ {\rm days}$ for $l=1$ in a non-rotating star.  }

\figcaption[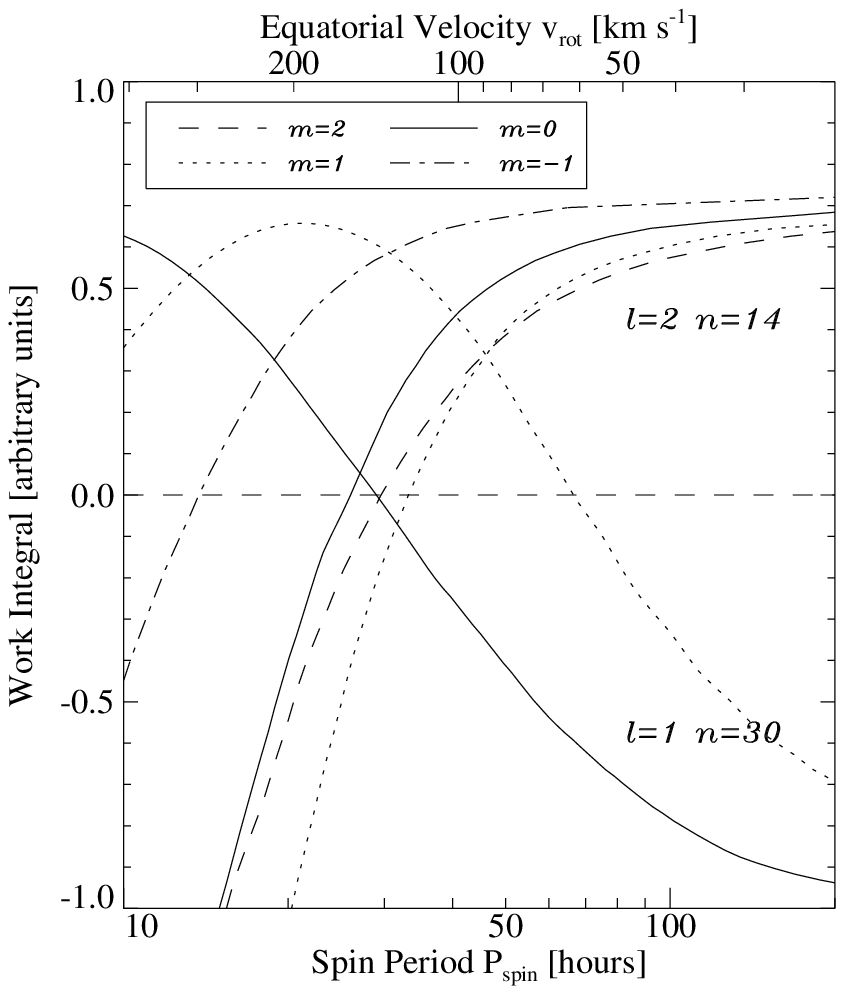]{ 
\label{fig:period}
Work integral $W$ (eq.~[\ref{eq:work-int}]) as a function of the spin
period $\Pspin$ (bottom axis) or equatorial rotational velocity
$\vrot$ (top axis) for our SPB model. Lines asymptoting to positive
$W$ for no rotation correspond to the $l=2$, $n=14$ mode shown in
Figure~\protect{\ref{fig:opacity-work}}b, while the lines asymptoting
to $W<0$ for no rotation correspond to the $l=1$, $n=30$ mode shown in
Figure~\protect{\ref{fig:opacity-work}}c. Different $m$ values are
marked on the plot, where $m>0$ modes are retrograde.}

\end{document}